# Intermodal group velocity engineering for broadband nonlinear optics

J. Demas,[1,2] L. Rishøj,[1] X. Liu,[1] G. Prabhakar,[1] and S. Ramachandran,[1,*]

[1]Boston University Department of Electrical Engineering, 8 St. Mary's St., Boston, MA 02115, USA
[2]Currently with The Rockefeller University, 1230 York Ave., New York, NY 10065, USA
*Corresponding author: sidr@bu.edu

**Interest in the nonlinear properties of multi-mode optical waveguides has seen a recent resurgence on account of the large dimensionality afforded by the platform. However, a perceived fundamental limitation of intermodal parametric interactions – that they are impractically narrowband – has yet to be solved. Here we show that by engineering the relative group velocity within the discrete spatial degree of freedom, we can tailor the phase matching bandwidth of intermodal parametric nonlinearities. We demonstrate group-velocity-tailored four-wave mixing between the $LP_{0,4}$ and $LP_{0,5}$ modes of a multi-mode fiber with unprecedented gain bandwidths (>60 nm at ~1550 nm). As evidence of the technological utility of this methodology, we seed this process to generate a high-peak-power wavelength-tunable fiber laser in the Ti:Sapphire wavelength regime. More generally, with the combination of intermodal interactions, which dramatically expand the phase matching degrees of freedom for nonlinear optics, and intermodal group velocity engineering, which enables tailoring the bandwidth of such interactions, we showcase a platform for nonlinear optics that can be broadband while being wavelength agnostic.**

## 1. Introduction

Multi-mode optical fibers provide the ability to guide light while encoding spatial, in addition to traditional temporal and spectral, information. Recently, there has been a resurgence of interest in using the spatial dimension inherent to this class of fibers for applications including imaging [1], high capacity telecommunications [2,3], and high power laser development [4,5]. Additionally, there has been considerable research into the nonlinear properties of these fibers for optical switching [6], parallel optical signal processing [7], frequency conversion [8—11], hybrid quantum entanglement [12], and a host of other complex nonlinear phenomena [13—15] unique to the multi-mode regime.

There is one main difficulty in harnessing the multi-mode space, however. As illustrated in the first demonstration of intermodal nonlinear optics by Stolen *et al.* in 1974 [16,17] – and repeatedly confirmed by subsequent experimental investigations to date [18—22] – nonlinear interactions between different spatial modes typically exhibit impractically narrow parametric gain bandwidths, limiting spectral or temporal tailoring of light, thereby constraining two crucial degrees of freedom in the process of exploiting the spatial degree of freedom. This narrow bandwidth obviates their utility for most known applications of parametric nonlinear interactions, such as multicasting classical communications signals [23], tailoring the joint-spectral amplitudes for quantum sources [24], or enabling ultrashort pulse nonlinear interactions, to name a few examples. This problem arises from the fact that nonlinear processes need to conserve momentum, or equivalently, phase-match. Transverse space in an optical fiber is discretized into modes rather than a continuous parameter, thus modes have discrete phase velocities, and accordingly only discrete (i.e. narrowband) phase-matched combinations are typically found.

Here, we demonstrate a solution to this problem by tailoring the group velocity of the $LP_{0,4}$ and $LP_{0,5}$ modes in a multi-mode optical fiber (Fig. 1a). By matching the group velocity of the interacting modes at their respective converted wavelengths, we can extend the phase-matched bandwidth for four-wave mixing in order to demonstrate, to the best of our knowledge, the first realization of broadband intermodal frequency conversion (63 nm at 1553 nm, 17 nm at 791 nm). Additionally, we seed this process (Fig. 1b) to generate a high-peak power (~10 kW) quasi-cw (~0.6 ns) source of wavelength-tunable radiation in the Ti:Sapphire wavelength band (786—795 nm). Thus, group-velocity-matched intermodal four-wave mixing unlocks a degree of freedom that decouples the interaction wavelengths in nonlinear optics from their respective bandwidths, while also providing for a power-scalable platform to achieve nonlinear frequency conversion.

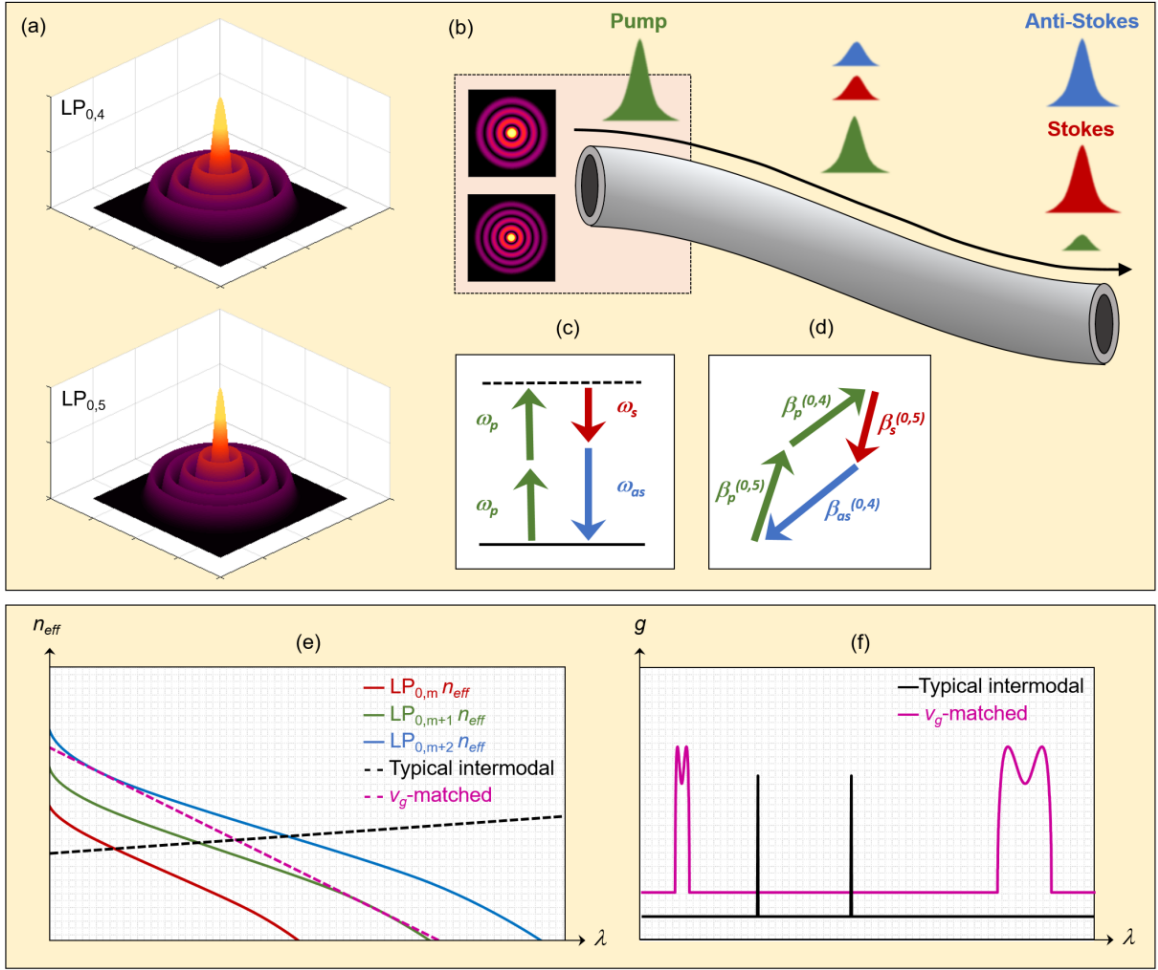

**Fig. 1** (a) Intensity profiles of the pump modes $LP_{0,4}$ and $LP_{0,5}$; (b) schematic representation of intermodal parametric frequency conversion of the pump (red pulse) to the anti-Stokes (blue pulse) and Stokes (green pulse) waves along the length of the fiber; schematic representation of (c) energy conservation and (d) phase matching for four-wave mixing; (e) phase matching in the effective refractive index picture: solutions exist where the straight, dashed lines intersect the $n_{eff}$ curves for different modes (solid lines); (f) parametric gain for typical intermodal processes (black line) and a group-velocity-tailored process (purple line), offset for clarity; the purple line in (e) is tangent to the $n_{eff}$ curves for an extended wavelength range leading to broadband parametric gain.

## 2. Intermodal phase matching

Four-wave mixing (FWM) is the nonlinear conversion of two "pump" photons to a higher frequency "anti-Stokes" photon and a lower frequency "Stokes" photon (Fig. 1). This process requires that the energy of the photons (proportional to their respective frequencies) and their momenta (proportional to the propagation constant $\beta$) be conserved (Figs. 1c and 1d). The parameter $\beta$ is related to the effective refractive index $n_{eff}$ by $\beta = 2\pi n_{eff}/\lambda$, thus the phase matching condition for FWM is given by

$$\Delta\beta = \frac{2\pi}{\lambda_1}n_{eff}^{(1)}(\lambda_1) + \frac{2\pi}{\lambda_2}n_{eff}^{(2)}(\lambda_2) - \frac{2\pi}{\lambda_3}n_{eff}^{(3)}(\lambda_3) - \frac{2\pi}{\lambda_4}n_{eff}^{(4)}(\lambda_4) \approx 0 \qquad (1)$$

where $\lambda$ is wavelength, the sub- and super- scripts '$j$' correspond to the wavelength and mode of the $j^{th}$ interacting wave, and the small offset due to self-phase modulation has been neglected [25]. It can be shown that for cases where the $n_{eff}$ of all four fields lie on a straight line (when plotted vs. wavelength), and the corresponding wavelengths conserve energy, the waves are necessarily phase-matched; allowing for a simple graphical representation of phase matching (see Supplement S2). This concept is illustrated in Fig. 1e, which shows two exemplary phase matching possibilities. In the typical intermodal case (black dashed line in Fig. 1e), the straight line intersects the $n_{eff}$ curves of three different modes (solid red, green, and blue lines), leading to phase matching. Here we see the utility of the multi-mode space, in that more modes provide more avenues for phase matching. However, these intersection points occur only for discrete sets of wavelengths, thus the resulting parametric gain (black line in Fig. 1f) is intrinsically narrowband.

The purple dashed line in Fig. 1e depicts the $n_{eff}$ phase matching picture for a group-velocity-tailored intermodal process. The pump laser is equally partitioned between two modes, thus the phase matching line lies between their respective $n_{eff}$ curves (solid blue and green lines Fig. 1e) at the pump wavelength. By adjusting the pump wavelength, we can precisely energy-match the pump to the anti-Stokes and Stokes wavelengths such that the $n_{eff}$ curves of the respective interacting modes are tangential to the phase matching line. This automatically yields broad gain bandwidths (purple line, Fig. 1f).

This condition is equivalent to the two modes being group-index-matched, and thus group-velocity-matched, at their respective anti-Stokes and Stokes wavelengths. This is evident by considering the phase matching problem in the frequency, rather than wavelength, domain, and recognizing that conservation of energy mandates that the frequency detuning $\Delta\omega$ for mode $j$ at the anti-Stokes frequency ($\omega_{as}$) corresponds to $-\Delta\omega$ for mode $k$ at the Stokes frequency ($\omega_s$), yielding (see detailed derivation in Supplement S2):

$$\Delta\beta = \Delta\omega \left( \frac{d\beta_k}{d\omega}\bigg|_{\omega_s} - \frac{d\beta_j}{d\omega}\bigg|_{\omega_{as}} \right) \quad (2)$$

For the unique case where the derivatives of $\beta$ for modes $j$ and $k$ evaluated at $\omega_{as}$ and $\omega_s$ are equal, the modes are phase-matched (i.e. $\Delta\beta = 0$) for arbitrary bandwidth (note that the higher order terms must be considered for very large $\Delta\omega$). Group velocity is given by $v_g = (d\beta/d\omega)^{-1}$, therefore matching the group velocities of the Stokes and anti-Stokes modes at their respective frequencies (wavelengths) maximizes the phase-matched bandwidth.

The relationship between group velocity and bandwidth here is analogous to phase matching in single-mode fibers (SMFs). Using the same straight line phase matching picture, the phase matching line for a single-mode system needs to intersect the same $n_{eff}$ curve at multiple wavelengths, which requires control over the wavelength-dependence of the group velocity, or the group velocity dispersion. Thus dispersion-engineered fibers, such as photonic crystal fibers, are crucial for single-mode nonlinear optics [26]. However, manipulating a single $n_{eff}$ curve limits the design space for phase matching – accordingly, there are two canonical parametric gain profiles in SMFs: (1) broadband gain at wavelengths adjacent to the pump laser, or (2), narrowband gain at wavelengths far from the pump [27—29]. By applying group-velocity engineering to multimode fibers, we can control parametric gain bandwidth while maintaining spatial degrees of freedom, allowing for an unconstrained design space for phase matching.

## 3. Experiment and results

Light from a home-built ~1 ns pulsed Ytterbium-doped fiber laser ($\lambda$ = 1048 nm) was coupled into a 2-m segment of step-index multi-mode test fiber (50-μm core diameter, numerical aperture = 0.22). We excite pump modes in the fiber using a binary phase plate programmed on a spatial light modulator per [30] (Fig. 2a). The phase plate structure is optimized to couple incident light into an equal superposition of the $LP_{0,4}$ and $LP_{0,5}$ modes at the pump wavelength. Simulations indicate that the mode purity is ~81% (see supplementary S1). The wavelength of the pump is chosen such that the group velocity (Fig. 2b) of the $LP_{0,4}$ mode at the anti-Stokes wavelength (7XX nm) is matched to that of the $LP_{0,5}$ mode at the Stokes wavelength (15XX nm). The corresponding simulated parametric gain curve shows broadband gain (Fig. 2c).

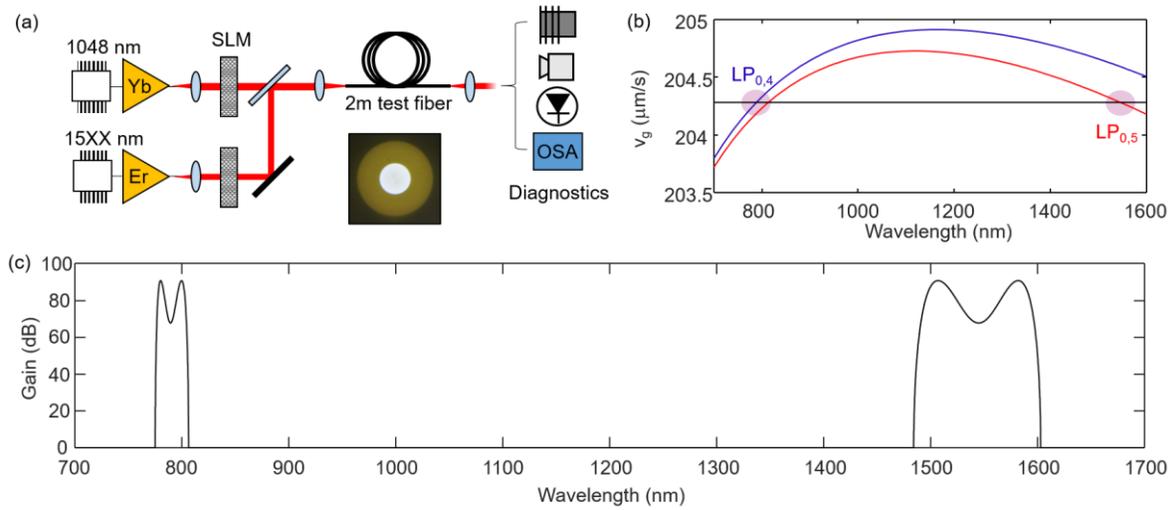

**Fig. 2** (a) Setup schematic for intermodal four-wave mixing experiments; SLM = spatial light modulator; OSA = optical spectrum analyzer; (b) simulated group velocity for the $LP_{0,4}$ and $LP_{0,5}$ modes; (c) simulated parametric gain for group-velocity-matched four-wave mixing pumped with $LP_{0,4}$ and $LP_{0,5}$.

When only the pump laser is coupled into the fiber, the resultant spontaneous parametric fluorescence spectrum reveals the spectral dependence of gain, and hence the phase matching wavelengths as well as bandwidths. Fig. 3a shows the parametric gain as a function of pump wavelength. Initially, there are two sets of gain peaks on the anti-Stokes and Stokes sides of the pump, respectively (purple curve, $\lambda = 1046.2$ nm). As the pump wavelength increases, the peaks move together and merge. For further increase in pump wavelength, the gain of the merged peak begins to decrease until it disappears. This can be understood with the intuition provided by Fig. 1e. Small changes in the pump wavelength change the energy matching requirements, and accordingly, change the slope of the phase matching line. For short pump wavelengths, the slope is steeper, and intersects the $n_{eff}$ curves twice on either side of the pump, resulting in two gain peaks. As the pump wavelength increases, the slope decreases, lying tangent to the $n_{eff}$ curves and merging the gain peaks into one broadband region. If the slope decreases further, it no longer intersects the curves; phase matching is not preserved, thus the gain disappears.

In order to maximize bandwidth, we operate at a pump wavelength of 1047.6 nm, where the line is slightly above the tangent point (see Fig. 1e for reference). The two gain peaks have not completely merged, resulting in the characteristic dip in the center of the gain bandwidth. The measured gain (Fig. 3b) exhibits broadband regions (10-dB bandwidths of 63 nm at 1553 nm, and 17 nm at 791 nm) in agreement with simulations (Fig. 2c). The measured image of the beam at the pump wavelength (inset Fig. 3b) does not resemble the profile of a single $LP_{0,m}$ mode and exhibits rings with poor visibility, a hallmark of the intended superposition of modes. In contrast, the measured mode images at the anti-Stokes and Stokes wavelengths (inset Fig. 3b) clearly correspond to the $LP_{0,4}$ and $LP_{0,5}$ modes, indicating wavelength conversion from the desired FWM process.

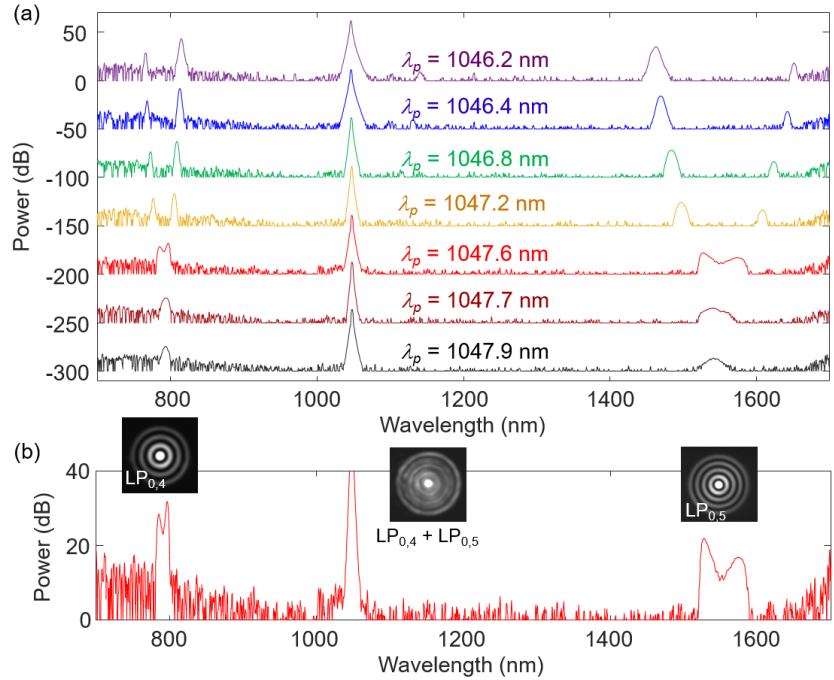

**Fig. 3** (a) Spontaneous four-wave mixing spectra as a function of pump wavelength; (b) Zoom-in of the bandwidth-optimized spontaneous spectrum ($\lambda_p = 1047.6$ nm) with mode images inset.

Next we introduce a low power erbium-doped fiber laser coupled into the $LP_{0,5}$ mode of the test fiber and tune its wavelength across the Stokes gain band (15XX nm). This second laser seeds the FWM process, leading to depletion of the pump and transfer of power to the Stokes and anti-Stokes waves. Fig. 4 shows measured output spectra for the system as a function of seed wavelength, indicating conversion of 1048 nm pump photons to the Ti:Sapphire spectral region (786-795 nm) while maintaining the narrow spectral linewidth of the seed laser.

The temporal profile of the output pulses are shown in Fig. 5a—5c for an example seed wavelength of 1545 nm. The center of the pump pulse undergoes depletion in the presence of the seed as the power is converted to the Stokes and anti-Stokes wavelengths (Fig. 5a). In total, $34 \pm 3$ % of the pump power is depleted, with further depletion limited by the leading and trailing edges of the pump pulse whose intensity is not sufficient to drive the nonlinear process (as is commonly observed in any nonlinear process with pump pulses that are not temporally rectangular in shape). As a result, the output pulse widths for the Stokes and anti-Stokes pulses are narrowed to ~0.6 ns (Fig. 5b and 5c). The conversion efficiency is $21 \pm 2$ % for the anti-Stokes wave, and $10 \pm 2$ % for the Stokes wave. The discrepancy between conversion efficiency at the anti-Stokes and Stokes wavelengths is due to the fact that FWM preserves photon number. Thus the anti-Stokes photons, which are roughly twice as energetic as the Stokes photons, constitute the majority of the output optical power. Normalizing for photon energy, the photon-to-photon transfer efficiency from the pump to the anti-Stokes and Stokes wavelengths are in agreement ($16 \pm 2$ % and $15 \pm 2$ %, respectively), as expected.

While the conversion efficiency is somewhat limited by the leading and trailing edges, the instantaneous pump depletion at the center of pulse is as high as 71% (purple markers in Fig. 5d). Because the peak of the pump pulse is efficiently transferred to the converted wavelengths, the output pulses have high peak powers of 11.2 kW for the anti-Stokes (Fig. 5b) and 6 kW for the Stokes (Fig. 5c) wavelengths, corresponding to ~48 dB of peak parametric gain for a seed wavelength of 1545 nm (brown markers in Fig. 5d). Instantaneous pump depletion remains high across the full seed tuning range (Fig. 5d) resulting in ~10 kW quasi-cw operation for all measured anti-Stokes pulses (Fig. 5e). At the edges of the tuning range, we see no significant degradation of the system performance (Fig. 5d and 5e), implying that operation across the full ~17 nm bandwidth predicted by the spontaneous parametric fluorescence (Fig. 3b) would have been possible were we not limited by the tunability of the seed source available in the laboratory.

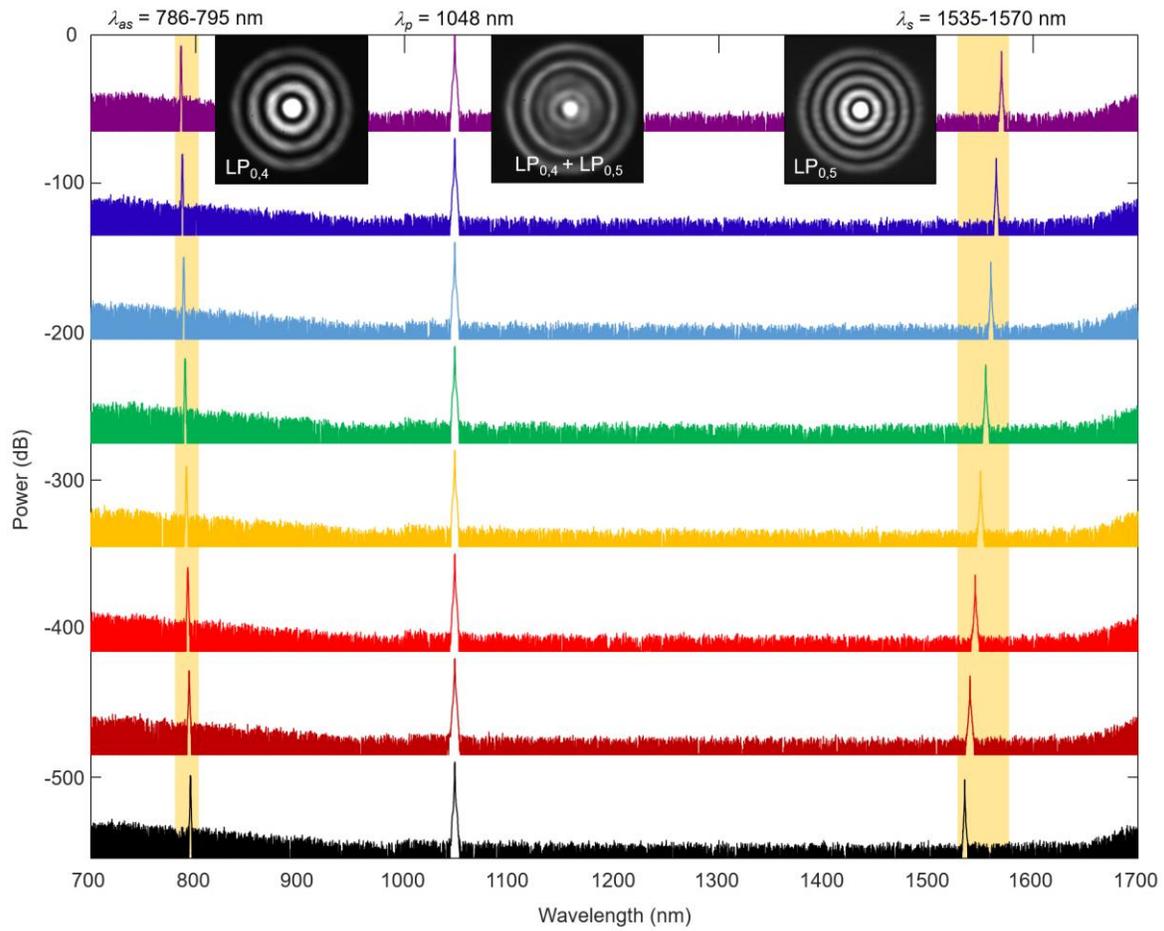

**Fig. 4** (a) Full experimental spectrum as a function of seed wavelength (1535-1570 nm) showing conversion to the Ti:Sapphire band (786-795 nm) with representative experimental mode images shown inset; each spectrum offset by 70 dB for clarity.

To the best of our knowledge, these results represent the first demonstration of a ~10 kW peak power wavelength tunable fiber source operating in the 7XX nm wavelength regime. The bandwidth supported by this system would enable conversion of pulses as short as ~40 fs. Hence, this represents the first all-fiber alternative to the ubiquitous Ti:Sapphire laser technology that is the mainstay of many optics and photonics applications today.

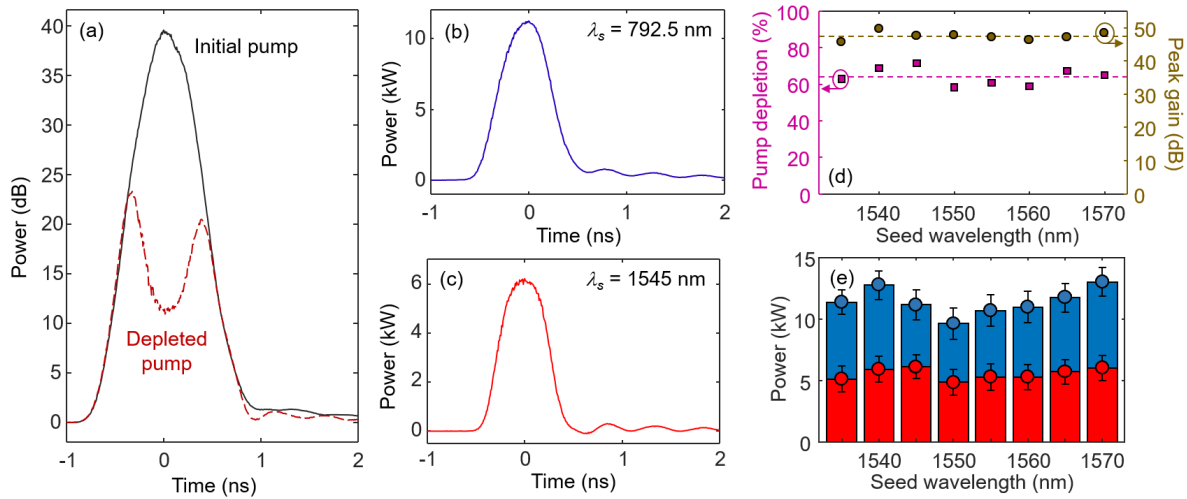

**Fig. 5** Experimentally measured pump profiles for the (a) pump (pump alone shown as a solid black line, pump combined with seed shown as a dotted purple line), (b) anti-Stokes, and (c) Stokes wavelengths for a 1545 nm wavelength seed; (d) peak pump depletion (purple markers, left axis) and peak Stokes gain (brown markers, right axis) as a function of seed wavelength; (e) peak power (anti-Stokes shown with blue markers, Stokes shown with red markers).

### 4. Discussion, summary and conclusions

We have shown that by employing group-velocity-matching between the $LP_{0,4}$ and $LP_{0,5}$ modes of a multi-mode fiber, we can upend the decades-old conception that intermodal parametric interactions must be impractically narrowband, and demonstrate a system that is wideband as well as broadband – a combination of attributes not afforded by single-mode nonlinear guided waves. This is possible because, unlike single mode waveguides, in systems with spatial diversity, phase matching (related to the propagation constant of light) is decoupled from bandwidth (related to the spectral gradient of the propagation). Indeed, the ubiquity of this concept is not restricted to the specific mode combinations we used in our experiments. Rather, as Table 1 illustrates, other processes with similar spectral and modal degeneracies can result in a wide variety of spectral ranges in which broadband intermodal nonlinear interactions can be obtained, simply by choice of mode order. More generally, we expect this phenomenon to hold true for modes with entirely different symmetries, and waveguide design may enable group-velocity-matched parametric nonlinearities in platforms other than fibers, such as on-chip waveguides.

**Table 1. Additional simulated broadband FWM processes**

| Pump modes | Pump wavelengths | Stokes gain bandwidth | Anti-Stokes gain bandwidth |
| --- | --- | --- | --- |
| $LP_{0,1} + LP_{0,2}$ | 1194 nm | 1519—1624 nm | 943—984 nm |
| $LP_{0,2} + LP_{0,3}$ | 1142 nm | 1548—1644 nm | 874—904 nm |
| $LP_{0,3} + LP_{0,4}$ | 1093 nm | 1533—1623 nm | 823—849 nm |
| $LP_{0,4} + LP_{0,5}$* | 1048 nm | 1502—1586 nm | 779—801 nm |
| $LP_{0,5} + LP_{0,6}$ | 1001 nm | 1469—1552 nm | 739—759 nm |

*corresponds to simulations of experiments described here*

Technologically, employing group velocity engineering to multi-mode nonlinear optics, reveals two benefits: (1) by enabling bandwidth-tailoring independent of phase matching constraints, it facilitates tailoring the spectral distribution of the nonlinear response agnostic of the wavelength of operation, thereby addressing applications such as multi-mode multicasting and factorable or non-factorable quantum state generation; and (2) by decoupling the phase matching condition and bandwidth from mode area, tunable frequency-converted sources become power-scalable in an all-fiber schematic. As such, intermodal group velocity engineering promises advances in controlling, exploiting, or managing optical nonlinearities in multi-mode optical systems, in analogy to the revolutionary advances in single-mode nonlinear optics made possible by dispersion engineering of photonic crystal designs.

**Materials and methods**

Further details regarding the materials and methods for these experiments are detailed in Supplement S1.

**Funding**

This work was supported by funding from the National Science Air Force Office of Scientific Research (AFOSR) BRI program (FA9550-14-1-0165) and the Office of Naval Research (ONR) (N00014-17-1-2519).

# Supplemental information for:

# Intermodal group velocity engineering for broadband nonlinear optics

J. Demas,[1,2] L. Rishøj,[1] X. Liu,[1] G. Prabhakar,[1] and S. Ramachandran,[1,*]

[1]Boston University Department of Electrical Engineering, 8 St. Mary's St., Boston, MA 02115, USA
[2]Currently with The Rockefeller University, 1230 York Ave., New York, NY 10065, USA
*Corresponding author: sidr@bu.edu

## S1. Materials and methods

Experiments were conducted with a home-built Ytterbium-doped fiber laser comprising light from a wavelength-tunable external cavity diode laser (Toptica DL-Pro 100) carved into ~1 ns pulses with an electro-optic modulator and amplified by three successive fiber amplification stages (peak power ~ 70 kW, $\lambda$ = 1040-1050 nm, 5 kHz repetition rate). The seed laser was constructed with a 15XX nm external cavity diode laser (Agilent HP 8168) with a single stage erbium-doped fiber amplifier (Pritel FA-32) and an acousto-optic modulator, yielding ~200 mW peak power, ~300 ns pulses tunable from 1535-1570 nm which are time-synchronized to the pump pulses. Further tuning of the seed was limited by the gain bandwidth of the fiber amplifier.

The test fiber was a 2-m segment of 50-μm core diameter step-index multimode fiber comprised of a pure silica core and fluorine-down-doped cladding with an index step of −0.017 (numerical aperture = 0.22, Thorlabs FG050LGA) which guides the first 10 $LP_{0,m}$ modes at the pump wavelength. Properties of the fiber (mode solutions, effective indices, group velocity, group velocity dispersion, etc.) were simulated using a home-built scalar eigenmode solver operating on a measurement of the fiber's refractive index profile (Interfiber Analysis IFA100). The length of the test fiber was determined by optimizing the pump depletion while cutting back the fiber length.

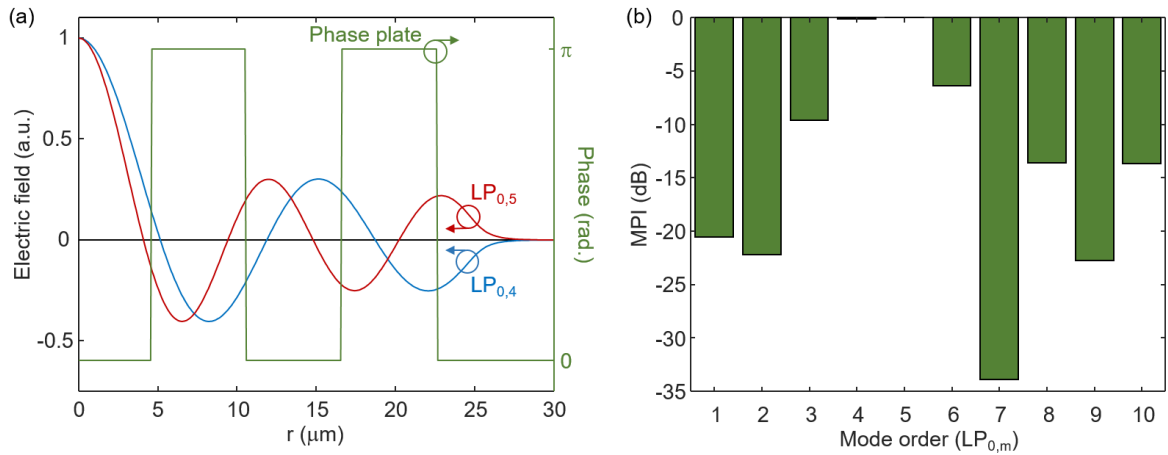

**Fig. S1** (a) Simulated electric field profiles of the $LP_{0,4}$ (blue line) and $LP_{0,5}$ (red line) modes plotted alongside the phase profile of a $LP_{0,5}$ binary phase plate stretched by ~12% (green line, right axis); (b) Simulated multi-path interference (MPI) as a function of mode order for coupling with the stretched binary phase plate.

Mode excitation was facilitated using two spatial light modulators (SLMs) in the seed and pump paths (Hammamatsu X10468-08 and X10468-07, respectively), each programmed with a binary phase plate designed to match the phase reversals of the $LP_{0,5}$ mode (as predicted by simulations of the modes in the test fiber). For each path, the Gaussian beam from the laser source was incident on the SLM where the binary phase was imparted. The two paths were combined using a dichroic mirror (Thorlabs DMSP1180) and finally each beam was imaged to the fiber facet using a telescope comprised of a lens programmed onto each SLM and a physical lens (f = 8 mm, Thorlabs C240TME-C) just before the fiber (following the method described in [32]).

For the pump path, the phase plate diameter was stretched larger than the optimal size for pure $LP_{0,5}$ coupling in order to simultaneously couple to both the $LP_{0,4}$ and $LP_{0,5}$ modes. As shown schematically in Fig. S1a, when the phase plate is appropriately stretched, the phase reversals occur roughly half-way between the nulls of the $LP_{0,4}$ and $LP_{0,5}$ modes, resulting in equal coupling to both modes. The phase plate size was optimized by coupling the pump laser to the test fiber while stretching the phase plate diameter in order to maximize the strength of the parametric fluorescence at the output of the fiber. The optimal phase plate size was stretched by ~12% relative to the size of a phase plate optimized for pure $LP_{0,5}$ coupling. Simulations indicate ~81% of the power launched into the fiber was coupled to the target mode. As shown in Fig. S1b, simulations predict that the majority of parasitic content is coupled to the $LP_{0,3}$ and $LP_{0,6}$ modes (where the multipath interference, or MPI, is defined as the normalized overlap integral between the incident field and each mode of the fiber). The coupling efficiency for the seed and pump beams was ~50-60%.

The output of the fiber was interrogated using an optical spectrum analyzer (ANDO AQ6317-B), a thermal power meter (Coherent PowerMax-USB), a silicon camera (Thorlabs DCC1645C), an InGaAs photo-detector (Thorlabs DET08C), and various near-infrared dielectric filters (Thorlabs, 10 nm FWHM).

## S2. Derivation of effective index phase matching line and group velocity dependence

*Derivation of phase matching line*

The phase matching condition for four-wave mixing expressed in terms of wavelength ($\lambda$) and effective refractive index ($n_{eff}$) is given by

$$\Delta\beta = \frac{2\pi}{\lambda_1} n_{eff}^{(1)}(\lambda_1) + \frac{2\pi}{\lambda_2} n_{eff}^{(2)}(\lambda_2) - \frac{2\pi}{\lambda_3} n_{eff}^{(3)}(\lambda_3) - \frac{2\pi}{\lambda_4} n_{eff}^{(4)}(\lambda_4) \approx 0 \tag{S1}$$

where the scripts correspond to the four interacting fields. We make the assumption that $n_{eff}$ for each of the interacting fields lies on a straight line such that $n_{eff}^{(k)}(\lambda_k) = m\lambda_k + b$, where $m$ and $b$ describe the slope and intercept of this "phase matching line," respectively. If we insert this definition for $n_{eff}$ into Eq. S1, the slope-dependent terms will cancel, yielding:

$$b\left(\frac{1}{\lambda^{(1)}} + \frac{1}{\lambda^{(2)}} - \frac{1}{\lambda^{(3)}} - \frac{1}{\lambda^{(4)}}\right) = 0 \tag{S2}$$

Given that the energy of a photon is proportional to its frequency $\omega$, and $\omega^{(k)} = 2\pi c/\lambda^{(k)}$ (where $c$ is the speed of light), this condition is exactly the energy-matching condition for four-wave mixing. Thus any four photons that lie on a straight line in effective index, at wavelengths that also uphold energy conservation, are necessarily phase matched.

*Derivation of group velocity dependence*

Consider a four-wave mixing interaction pumped between the $LP_{0,j}$ and $LP_{0,k}$ modes, with a Stokes field in the $LP_{0,k}$ mode and anti-Stokes field in the $LP_{0,j}$ mode. We assume that these fields are nominally phase matched at frequencies $\omega_p$, $\omega_s$, and $\omega_{as}$ (where '*p*' corresponds to pump, '*s*' corresponds to Stokes, and '*as*' corresponds to anti-Stokes) such that:

$$\beta_j(\omega_p) + \beta_k(\omega_p) - \beta_j(\omega_{as}) - \beta_k(\omega_s) = 0 \tag{S3}$$

where $\beta = \omega n_{eff}(\omega)/c$. We also assume energy conservation is upheld such that $2\omega_p - \omega_s - \omega_{as} = 0$. To determine the frequency dependence of the phase matching condition, we define the phase mismatch $\Delta\beta$ for a small frequency detuning $\Delta\omega$ from this initially phase-matched solution:

$$\Delta\beta = \beta_j(\omega_p) + \beta_k(\omega_p) - \beta_j(\omega_{as} + \Delta\omega) - \beta_k(\omega_s - \Delta\omega) \tag{S4}$$

Note that $\Delta\omega$ is opposite in sign for the anti-Stokes and Stokes terms in order to conserve energy. The phase matching bandwidth (and thus parametric gain bandwidth) corresponds to the frequency range $\Delta\omega$ for which $\Delta\beta \approx 0$. In order to evaluate $\Delta\beta$, we Taylor expand the Stokes and anti-Stokes terms to first order yielding:

$$\Delta\beta = \beta_j(\omega_p) + \beta_k(\omega_p) - \left(\beta_j(\omega_{as}) + \Delta\omega \left.\frac{d\beta_j}{d\omega}\right|_{\omega_{as}}\right) - \left(\beta_k(\omega_s) - \Delta\omega \left.\frac{d\beta_k}{d\omega}\right|_{\omega_s}\right) \tag{S5}$$

Recognizing that the zeroth order terms will cancel, as the fields are nominally phase-matched (Eq. S3), the relation simplifies to the following:

$$\Delta\beta = \Delta\omega \left( \frac{d\beta_k}{d\omega}\bigg|_{\omega_s} - \frac{d\beta_j}{d\omega}\bigg|_{\omega_{as}} \right) \tag{S6}$$

The first derivative of $\beta$ with respect to $\omega$ is the reciprocal of group velocity ($v_g = [d\beta/d\omega]^{-1}$), thus this expression indicates that the phase-matched bandwidth is related to the group velocity mismatch between the interacting modes at the Stokes and anti-Stokes frequencies. For the unique case where the Stokes and anti-Stokes modes are group-velocity-matched, $\Delta\beta$ is identically zero independent of $\Delta\omega$, and phase matching is upheld for a large, but indefinite bandwidth. In this instance, the higher order terms must be considered, specifically the group velocity dispersion (GVD = $d^2\beta/d\omega^2$). Assuming group index matching (and thus cancellation of the first order terms), we can expand Eq. S5 to 2nd order and arrive at:

$$\Delta\beta = \Delta\omega^2 \left( \frac{d^2\beta_j}{d\omega^2}\bigg|_{\omega_{as}} + \frac{d^2\beta_k}{d\omega^2}\bigg|_{\omega_s} \right) \tag{S7}$$

In our case, the group velocity dispersion values for the Stokes and anti-Stokes modes are −56.2 and 34 ps$^2$/km respectively, and thus far from canceling in Eq. S7. Nonetheless, this mismatch still allows for phase matching bandwidths of ~49 THz ($\Delta\lambda = 63$ nm at 1553 nm). Through group velocity dispersion engineering, it may be possible to make the dispersion of the Stokes and anti-Stokes waves equal in magnitude but opposite in sign and extend the phase-matched bandwidth even further.